\documentclass[runningheads]{llncs}
\usepackage[T1]{fontenc}
\usepackage{graphicx,subcaption}
\usepackage{caption, enumitem, url}
\usepackage{natbib,xcolor}
\usepackage{hyperref}       
\usepackage{amsmath,amssymb}
\usepackage[mathlines]{lineno} 
\begin{document}
\title{Approximating the universal thermal climate index using sparse regression with orthogonal polynomials}
\titlerunning{Approx. the UTCI using sparse regression with orthogonal polynomials}
%
\author{Sabin Roman \inst{1} \and Ljup{\v{c}}o Todorovski\inst{2, 1}\and
Sa{\v{s}}o D{\v{z}}eroski \inst{1}
\and Gregor Skok\inst{2}
}
%
%
\institute{Department of Knowledge Technologies, Jo\v{z}ef Stefan Institute, Jamova 39, 1000 Ljubljana, Slovenia\\
\and
Faculty of Mathematics and Physics, University of Ljubljana, Jadranska ulica 19, 1000 Ljubljana, Slovenia\\
}
\maketitle              
\begin{center}
Correspondence: Sabin Roman (sabin.roman@ijs.si)
\end{center}
\begin{abstract}
The Universal Thermal Climate Index (UTCI) is a measure of thermal comfort that quantifies how humans experience environmental conditions. Due to its robustness and versatility as a bioclimatic indicator, it has been extensively employed across a wide range of studies in bioclimatology and is increasingly used as an operational measure of outdoor thermal comfort. At the same time, calculating the UTCI value from the relevant environmental parameters is nominally not straightforward, which is why using a 6th-degree polynomial approximation has become the standard way to calculate UTCI values. At the same time, although it is computationally efficient, the error of this polynomial approximation can be substantial. The goal of this study was to develop an improved version of the polynomial approximation -- one that retains comparable computational efficiency but is more robust in terms of numerical stability and substantially more accurate, particularly in reducing the frequency of larger errors. This goal was successfully achieved using sparse orthogonal regression, namely sparse regression with an orthogonal polynomial basis, which not only substantially reduces the average errors (i.e., the mean error, the mean absolute error, and the root mean square error) but also drastically reduces the frequency of large errors. By leveraging Legendre polynomial bases, approximation models could be constructed that efficiently populate a Pareto front of accuracy versus complexity and exhibit stable, hierarchical coefficient structures across varying model capacities. Training the new approximation models over only 20\% of the data, with the testing performed over the remaining 80\%, highlights successful generalization, with the results also being robust under bootstrapping. The decomposition effectively approximates the UTCI as a Fourier-like expansion in an orthogonal basis, yielding results near the theoretical optimum in the $L_{2}$ (least squares) sense.

\end{abstract}
\section{Introduction}

The Universal Thermal Climate Index (UTCI) is a measure of thermal comfort that quantifies how humans experience environmental conditions. It is derived from an advanced thermo-physiological model \citep{Pappenberger2015} and expressed in units of temperature. The index accounts for multiple factors, including air temperature, humidity, wind speed, radiation, and clothing insulation \citep{Brode2012}. A notable advantage of the UTCI compared to many other bioclimatic indices is its ability to represent thermal conditions in terms that are applicable to human strain under a wide range of climatic conditions (e.g., for both hot and cold conditions, \cite{Blazejczyk2012}). Based on the UTCI value, the environmental conditions can be classified into one of the ten thermal stress categories \citep{Brode2012}, ranging from Extreme heat stress ($UTCI > 43$$^{\circ}$C) to Extreme cold stress ($UTCI < -40$$^{\circ}$C).

Owing to its robustness and versatility as a bioclimatic indicator, the UTCI has been extensively employed across a wide range of studies in bioclimatology and related scientific disciplines. Its applications encompass diverse research areas, including the assessment of regional and local bioclimate characteristics, the study of urban bioclimate, recreation, tourism, and sports, epidemiological and health-related research, as well as the assessment and forecasting of bioclimatic changes \citep{utcistudes}. The UTCI has also seen growing adoption across numerous countries as a standardized measure of outdoor thermal comfort and is increasingly integrated into routine operational meteorological forecasts. For example, within Europe, UTCI is used operationally in the Czech Republic, Italy, Poland, Portugal, and Slovenia \citep{utciaplic,Kuzmanovi2024}.

At the same time, calculating the UTCI value from the relevant environmental parameters is nominally not straightforward. Namely, the UTCI is based on the Fiala multi-node model of human thermoregulation \citep{multinode}. However, running the complete Fiala model is computationally expensive and requires expert knowledge to operate the complex simulation software \citep{Brode2012}. This is the reason the authors of \cite{Brode2012} provided two simplified approximate procedures for calculating the UTCI values that could be used in operational settings. The first approximation is based on a 4-dimensional look-up table of $104\,643$ accurate pre-calculated UTCI values that cover a wide range of relevant combinations of the meteorological parameters. Using this look-up table, interpolation from nearby data points can be used to determine approximate UTCI values for intermediate values of meteorological parameters. The second approximation is based on a 6th-degree regression polynomial with 210 coefficients. 

Each approximation has its benefits and weaknesses. The look-up table approach is more accurate, but storing the tabulated values and searching for neighboring datapoints poses challenges to the implementation of this algorithm, while also resulting in a longer execution time compared to the other approach \citep{Brode2021}. In contrast, the polynomial approximation is less accurate, but computationally faster and substantially easier to implement in various programming languages and computational environments, as it relies on only the most common, primitive mathematical operators and does not require storing the tabulated values. At the same time, the motivation for improving the polynomial approximation is not simply a matter of storage, since the size of the look-up table is modest in modern computational settings. Rather, an improved polynomial approximation remains attractive for several practical reasons:
\begin{enumerate}
    \item[(i)] It is fully self-contained and does not depend on external tabulated data, which facilitates reproducibility and makes redistribution and integration into open-source software and operational tools more straightforward;
    \item[(ii)] It is computationally more efficient than look-up-table-based interpolation, which has been reported to be slower by roughly three orders of magnitude \citep{Brode2012}, an important consideration in large-scale applications such as numerical weather prediction and climate reanalysis;
    \item[(iii)] It is simpler to implement and port across programming languages and computational environments, including constrained, embedded, or legacy systems, because it requires only basic arithmetic operations and avoids the additional logic needed for multidimensional interpolation, data handling, and neighborhood search;
    \item[(iv)] It provides a direct, continuous, and analytically defined mapping over the domain of validity, whereas the look-up table still requires interpolation, and in some cases extrapolation, for environmental states not explicitly represented in the tabulated values;
    \item[(v)] Its predictive behavior on unseen data can be assessed directly through a train--test evaluation framework; in the present case, training on 20\% of the dataset and testing on the remaining 80\% still yields very good predictive performance, indicating strong generalization.
\end{enumerate}
For these reasons, the polynomial approximation is best viewed not as a universal replacement for the look-up-table approach, but as a complementary alternative that is particularly useful in applications where speed, portability, reproducibility, and ease of deployment are important.

Due to its simplicity and computational efficiency, the polynomial approximation has become the standard way of calculating the UTCI values. It has been incorporated into various bioclimatic software packages and libraries (e.g., the Bioklima software, \cite{BioKlimasite}, the Thermofeel Python library, \cite{Brimicombe2022}, and the pyThermalComfort Python library \cite{Tartarini2020}), as well as numerical weather prediction and reanalysis systems (e.g., the ALADIN model, \cite{aladinTermonia}, and the ERA5 reanalysis, \cite{ECMWFRayman}). At the same time, the error of the polynomial approximation can be substantial. For example, when evaluated on the aforementioned look-up table of accurate UTCI values, the root-mean-square-error is about 1.1$^{\circ}$C while the frequency of absolute errors larger than 2$^{\circ}$C is about 8\%, and the frequency of errors larger than 3$^{\circ}$C is about 2\%. This is problematic since an error of a few degrees Celsius can increase the likelihood of misclassification of the thermal stress category, some of which span only a 6$^{\circ}$C interval. 

The goal of this study is to develop an improved version of the polynomial approximation -- one that has comparable computational complexity to the existing approximation but is more robust in terms of numerical stability and substantially more accurate, particularly in reducing the frequency of larger errors. To achieve this goal, symbolic and sparse regression techniques are used as tools for interpretable and efficient function approximation. We fit the UTCI offset using sparse regression on an orthogonal Legendre polynomial basis. To emphasize this key feature and distinguish it from standard sparse regression on monomials, we refer to this approach as sparse orthogonal regression.

\begin{table}[t]
\centering
\begin{tabular}{c|c|c|c}
\textbf{Variable} & \textbf{Description} & \textbf{Valid} & \textbf{Normalized} \\ \textbf{name} & \textbf{} & \textbf{Range} & \textbf{range} \\ \hline
$\mathit{Ta}$ & Air temperature & $-50~$to$~+50$ $^{\circ}$C & $[-1, 1]$ \\
$\mathit{va}$ & Wind speed at 10 m & $0.5~$to$~30.3$ m/s & $[-1, 1]$ \\
$\mathit{Tr}-\mathit{Ta}$ & Mean Radiant–air temperature difference & $-30~$to$~+70$ $^{\circ}$C & $[-1, 1]$ \\
$\mathit{rH}$ & Relative humidity & $5~$to$~100$ \% & $[-1, 1]$ \\
$\mathit{pa}$ & Water vapour pressure & $0~$to$~5$ kPa & Not used \\
\end{tabular}%
\caption{Description of variables used in this study, following \cite{Brode2012}. The normalized ranges map each variable to $[-1,1]$, with respect to the interval of validity, suitable for use with Legendre polynomial bases. Although water vapor pressure ($\mathit{pa}$) is not used directly as an input for the new approximation, it can be computed from air temperature ($\mathit{Ta}$) and relative humidity ($\mathit{rH}$), and its effect is therefore accounted for through the inclusion of $\mathit{rH}$.}
\label{tab:vars}
\vspace{-5mm}
\end{table}

We also note that the aim was not to derive an approximation that was as accurate as possible. For example, a sufficiently complex neural-network-based model would likely provide more accurate estimates of the UTCI values. However, such a model would also require the use of machine-learning libraries, as well as suitable Graphics Processing Units, to function efficiently. This means that its implementation in various programming languages and computational environments would be substantially more difficult. On the other hand, replacing an existing polynomial approximation with a new one is fairly straightforward, meaning that implementing the new approximation into existing biclimatic software packages/libraries and numerical weather prediction systems would be relatively easy. 

\section{Methods}

Formally, the UTCI is defined as \citep{Brode2012}

\begin{equation}
\mathit{UTCI} = \mathit{Ta} +  \text{Offset}(\mathit{Ta}, \mathit{va}, \mathit{Tr}, \mathit{rH}\text{ or }\mathit{pa}),
\label{utciapx}
\end{equation}
where $\mathit{Ta}$ is the air temperature and the Offset is the physiologically equivalent temperature difference, representing how other environmental factors modify the effect of the thermal stress on the human body. The Offset function represents the deviation of the UTCI from the actual air temperature and depends on $\mathit{Ta}$, wind speed at 10~m ($\mathit{va}$), mean radiant temperature ($\mathit{Tr}$), which accounts for the effect of all incoming radiation, and humidity, which can be represented by either relative humidity ($\mathit{rH}$) or water vapour pressure ($\mathit{pa}$).

The dataset provided by \cite{Brode2012} contains accurate values of the Offset function covering a wide range of environmental states. The variables and their ranges are included in Table~\ref{tab:vars}. The intervals of the environmental variables also represent the domain where the sixth-degree polynomial regression approximation is considered valid \citep{Brode2012}. Using the approximation for conditions outside of these intervals can lead to large errors and unrealistic values of the Offset function and should be avoided \citep{Brode2021}.

\begin{figure}[t]
\captionsetup[subfigure]{labelformat=empty}
  \centering
  \subfloat[][(a)]{\includegraphics[width=.48\textwidth]{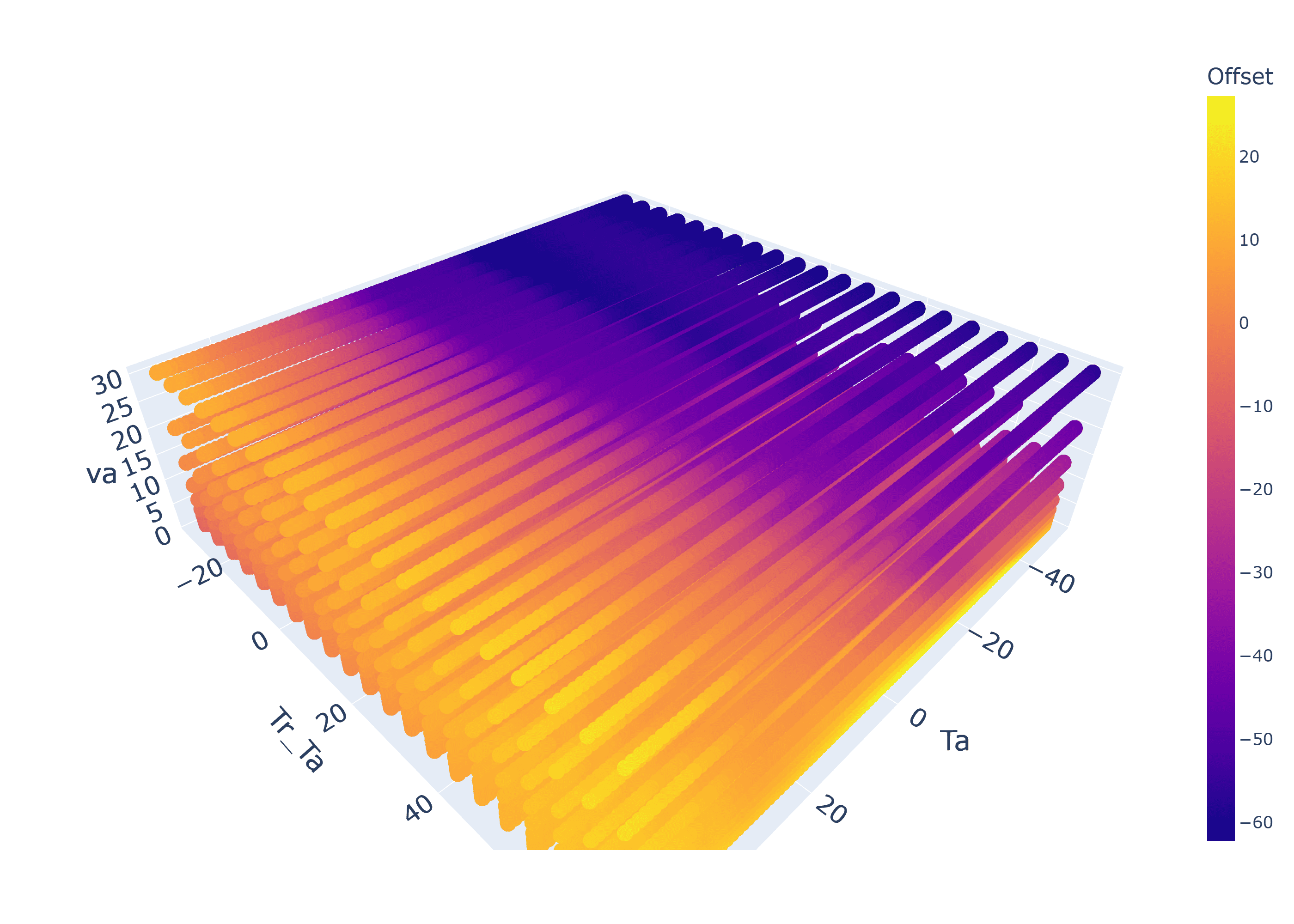}}\quad
  \subfloat[][(b)]{\includegraphics[width=.48\textwidth]{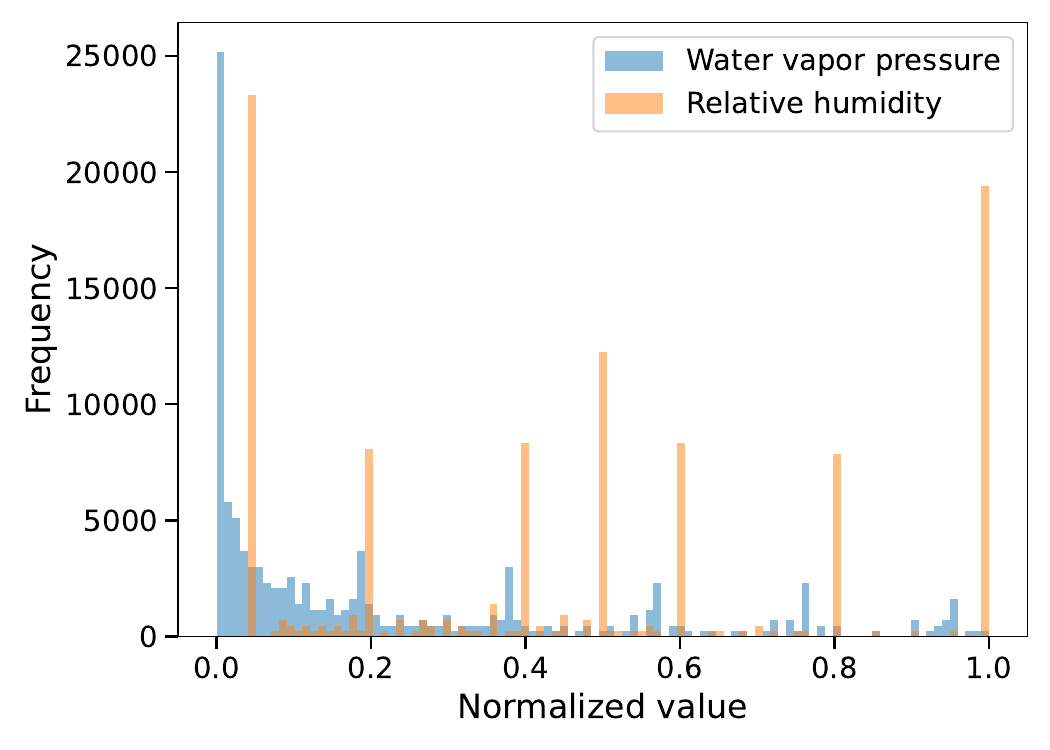}}
    \caption{(a) 3D plot of UTCI Offset \cite{Brode2012} at 5\% relative humidity, showing how wind speed ($\mathit{va}$), air temperature ($\mathit{Ta}$), and mean radiant temperature difference ($\mathit{Tr}-\mathit{Ta}$) combine to influence thermal stress. Color indicates the UTCI Offset magnitude across these environmental dimensions. (b)  The different distributions of the water vapor pressure and relative humidity in the computed Offset dataset \cite{Brode2012}. The water vapor pressure is strongly peaked at zero, while the relative humidity is uniform across its range.}
    \label{fig:2}
\vspace{-5mm}
\end{figure}

In Fig.~\ref{fig:2}(a) we see how the UTCI Offset varies along the different environmental variables. Instead of the humidity ($\mathit{rH}$), the water vapor pressure ($\mathit{pa}$) can be used which is a nonlinear function of $\mathit{rH}$ and the air temperature ($\mathit{Ta}$). However, the variables have different distribution, see Fig.~\ref{fig:2}(b), which impacts the extent that approximations of UTCI can generalize, discussed below.

Equation discovery aims to learn interpretable mathematical expressions, either differential or algebraic equations, from measurements of the variables of a given observed system \citep{todorovski1997declarative}. Positioned at the intersection of symbolic machine learning and system identification, it is becoming increasingly relevant in environmental and climate science, where data-driven yet transparent models are essential \citep{steinmann2025scenario,roman2025maximum}. Traditional modeling approaches rely on expert-derived formulations \citep{roman2021historical,roman2023theories,roman2022master,roman2023global,roman2019growth}, but the growing complexity and volume of climate data call for automated alternatives. Symbolic regression, which iteratively combines mathematical operators and variables to fit data, forms the core of equation discovery \citep{bridewell2005reducing,todorovski2006integrating,dvzeroski2007computational}. Most methods employ evolutionary or other (e.g., enumerative) search strategies to explore the space of candidate equations \citep{tanevski2016learning,tanevski2020combinatorial,mevznar2023efficient}.

Recent advances integrate probabilistic grammars to incorporate prior knowledge and constrain the search to physically meaningful expressions \citep{brence2020probabilistic,brence2023dimensionally, omejc2024probabilistic}. This structured approach improves both model interpretability and search efficiency, especially in domains governed by established scientific principles. Equation discovery has been applied to various environmental systems \citep{atanasova2006computational,atanasova2011automated,atanasova2008application,atanasova2006automated}, including ecosystem dynamics \citep{jeraj2006application,vcerepnalkoski2012influence,simidjievski2015learning,simidjievski2016modeling,tanevski2016process}. In these settings, it can match or even surpass expert-built models while simultaneously revealing new relationships \citep{todorovski2001theory,todorovski1998modelling}. Its ability to generate compact, interpretable, and physically plausible models makes it especially suitable for climate applications, where model transparency and adherence to physical principles are vital.

\begin{figure}[t]
\captionsetup[subfigure]{labelformat=empty}
  \centering
  \subfloat[][(a)]{\includegraphics[width=.48\textwidth]{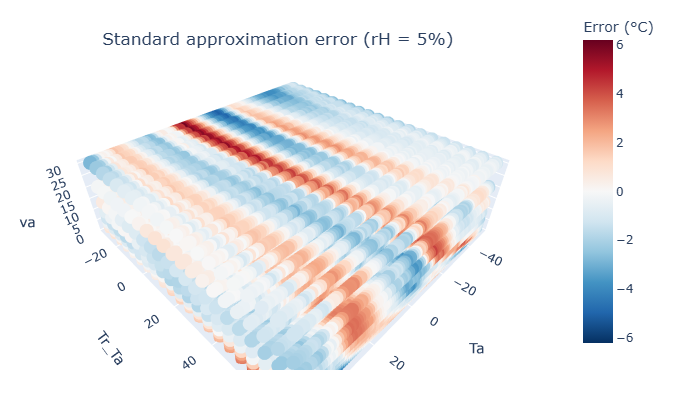}}\quad
  \subfloat[][(b)]{\includegraphics[width=.48\textwidth]{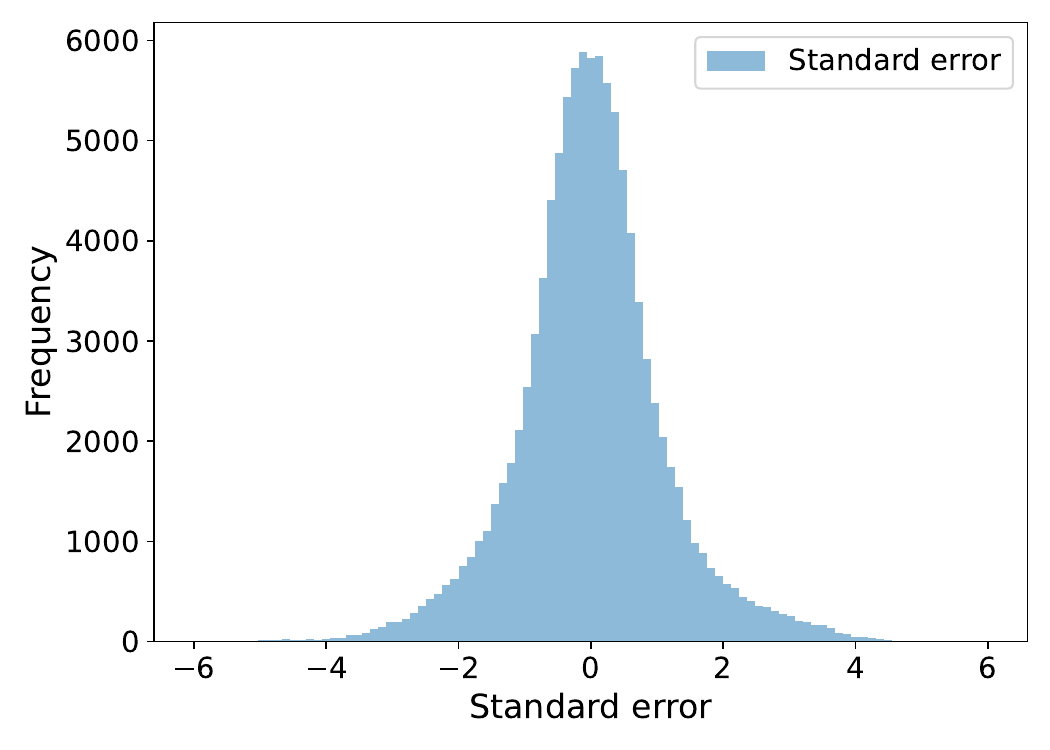}}
    \caption{The error of the standard polynomial UTCI approximation \citep{Brode2012} for relative humidity of 5\%. (a) The difference between the standard UTCI approximation and the accurate values of the Offset function. (b) Histogram of the differences showing a normal distribution centered at zero.}
    \label{fig:3}
\vspace{-5mm}
\end{figure}

As already mentioned, the errors of the sixth-degree regression polynomial from \cite{Brode2012} can be substantial. Fig.~\ref{fig:3}(a) shows the approximation error at 5\% relative humidity, while \ref{fig:3}(b) displays a histogram of the errors, revealing a normal distribution centered at zero, indicating minimal bias. We aim to improve upon this standard approximation using equation discovery and sparse regression methods by utilizing the accurate Offset dataset provided by \cite{Brode2012}.

Sparse machine learning models aim to construct parsimonious predictive functions by enforcing zero-valued coefficients in high-dimensional parameter spaces, thereby performing implicit feature selection \citep{brunton2016discovering}. This sparsity promotes interpretability, reduces overfitting, and improves computational tractability, especially when the number of candidate predictors is large or when strong correlations exist among inputs. Sparse regression \citep{brunton2016discovering}, a key instantiation of this paradigm, extends linear regression with an $L_1$-norm regularization term—most notably in the Lasso \citep[Least Absolute Shrinkage and Selection Operator,][]{reid2016study} —to penalize unnecessary parameters and induce a compact representation.

In this work, we employ sparse regression to identify compact, interpretable models of the UTCI, emphasizing its suitability for high-dimensional input spaces with redundant or weakly relevant features. While sparse modeling is well-established in statistical learning, its application to orthogonal polynomial bases-particularly in the context of bioclimatic indices—remains unexplored. By leveraging the structure of orthogonal polynomials, we obtain improved numerical stability and additive expansions that facilitate coefficient interpretability. To our knowledge, this is the first application of sparse regression using orthogonal bases to approximate the UTCI, addressing both predictive accuracy and model parsimony. Our results show that this approach surpasses the standard sixth-degree polynomial approximation in both accuracy and efficiency.

\section{Results and discussion}

Table~\ref{tab:1} presents a detailed comparison of model performance across a range of polynomial degrees for both standard (non-sparse) linear regression and sparse regression techniques, evaluated in the context of approximating the UTCI. The standard approximation \citep{Brode2012} is a sixth-degree regression polynomial model with four variables, consisting of 210 terms and achieving a root mean squared loss of $1.12^\circ$C. This serves as the benchmark to be matched or improved upon. It is important to note that the standard approximation does not directly employ the relative humidity ($\mathit{rH}$), but the water vapor pressure ($\mathit{pa}$), which can be derived from the relative humidity ($\mathit{rH}$) and air temperature ($\mathit{Ta}$). As we noted above, in the dataset, the relative humidity is well represented across its entire range, see Fig.~\ref{fig:2}(b), while the water vapor pressure is strongly peaked close to zero. Optimization employing the water vapor pressure ($\mathit{pa}$) as an independent variable (instead of $\mathit{rH}$) is thus poorly conditioned and leads to instability in the regression coefficients, both in simple and sparse regression. While using the pa (instead of $\mathit{rH}$) can achieve better accuracy (lower loss), it comes at the price of losing parameter consistency across optimizations with different polynomial degrees. For this reason, we report our results employing the relative humidity ($\mathit{rH}$) instead of the water vapor pressure ($\mathit{pa}$), see Table~\ref{tab:vars}.

\setlength\tabcolsep{1em}
\begin{table}[t]
\centering
\begin{tabular}{c|ccccccc}
Method &
  \multicolumn{7}{c}{} \\ \hline
Standard &
  \multicolumn{7}{c}{\begin{tabular}[c]{@{}c@{}}1.12\\ (210)\end{tabular}} \\ \hline
 &
  \multicolumn{7}{c}{Polynomial degree} \\ \hline
 &
  \multicolumn{1}{c|}{4th} &
  \multicolumn{1}{c|}{6th} &
  \multicolumn{1}{c|}{8th} &
  \multicolumn{1}{c|}{10th} &
  \multicolumn{1}{c|}{12th} &
  \multicolumn{1}{c|}{14th} &
  16th \\ \hline
\begin{tabular}[c]{@{}c@{}}Linear\\ regression\end{tabular} &
  \multicolumn{1}{c|}{\begin{tabular}[c]{@{}c@{}}2.1\\ (70)\end{tabular}} &
  \multicolumn{1}{c|}{\begin{tabular}[c]{@{}c@{}}1.3\\ (210)\end{tabular}} &
  \multicolumn{1}{c|}{\begin{tabular}[c]{@{}c@{}}0.92\\ (495)\end{tabular}} &
  \multicolumn{1}{c|}{\begin{tabular}[c]{@{}c@{}}Train: 0.67\\ Test: 0.71\\ (1001)\end{tabular}} &
  \multicolumn{1}{c|}{\begin{tabular}[c]{@{}c@{}} 0.54\\ 0.62\\ (1820)\end{tabular}} &
  \multicolumn{1}{c|}{\begin{tabular}[c]{@{}c@{}}0.44\\ 0.66\\ (3060)\end{tabular}} &
  \begin{tabular}[c]{@{}c@{}}0.36\\ 1.74\\ (4845)\end{tabular} \\ \hline
\begin{tabular}[c]{@{}c@{}}Sparse\\orthogonal\\ regression\end{tabular} &
  \multicolumn{1}{c|}{\begin{tabular}[c]{@{}c@{}}2.1\\ (65)\end{tabular}} &
  \multicolumn{1}{c|}{\begin{tabular}[c]{@{}c@{}}1.38\\ (124)\end{tabular}} &
  \multicolumn{1}{c|}{\begin{tabular}[c]{@{}c@{}}1.03\\ (176)\end{tabular}} &
  \multicolumn{1}{c|}{\begin{tabular}[c]{@{}c@{}}\textbf{0.88}\\ (209)\end{tabular}} &
  \multicolumn{1}{c|}{\begin{tabular}[c]{@{}c@{}}0.69\\ (355)\end{tabular}} &
  \multicolumn{1}{c|}{\begin{tabular}[c]{@{}c@{}}0.63\\ (400)\end{tabular}} &
  \begin{tabular}[c]{@{}c@{}}0.6\\ (424)\end{tabular}
\end{tabular}
\caption{Root mean squared train loss [$^{\circ}$C], test loss [$^{\circ}$C] and the number of parameters (shown in parenthesis) in approximating the UTCI Offset. The baseline reference, labeled as ``Standard,'' corresponds to the sixth-degree regression polynomial model with four variables \citep{Brode2012}. Unless otherwise stated the test loss equals the train loss. Where two loss values are reported (train loss on the top and test loss below), they indicate a notable train-test discrepancy, typically suggesting overfitting. Training is done with 20\% of the data and testing is performed with 80\%. Results are robust under bootstrapping.}
\label{tab:1}
\vspace{-5mm}
\end{table}

The regression methods are applied to polynomial basis expansions of increasing degree, evaluated on the basis of root mean squared test loss and number of active parameters. Unlike many studies in the literature where models are trained on the majority of the data and evaluated on a relatively small test set, our approach inverts this paradigm: training is conducted on only 20\% of the available data, while performance is assessed on the remaining 80\%. Despite this stringent evaluation setting, the models achieve comparable performance on both training and test sets, underscoring their strong generalization capabilities. This performance stability is further validated through bootstrapping, which reveals minimal variance in both loss metrics and selected features across resampled datasets. The reported performance metrics—such as train/test loss and number of parameters—remain stable when the model training and evaluation process is repeated on multiple random re-samplings (bootstrapped subsets) of the data. This suggests that the results are not sensitive to specific data splits and that the models generalize well across different subsets of the dataset, indicating reliability and consistency in the reported findings. These findings demonstrate the robustness and reliability of the proposed framework.

\begin{figure}[t]
\includegraphics[width=\textwidth]{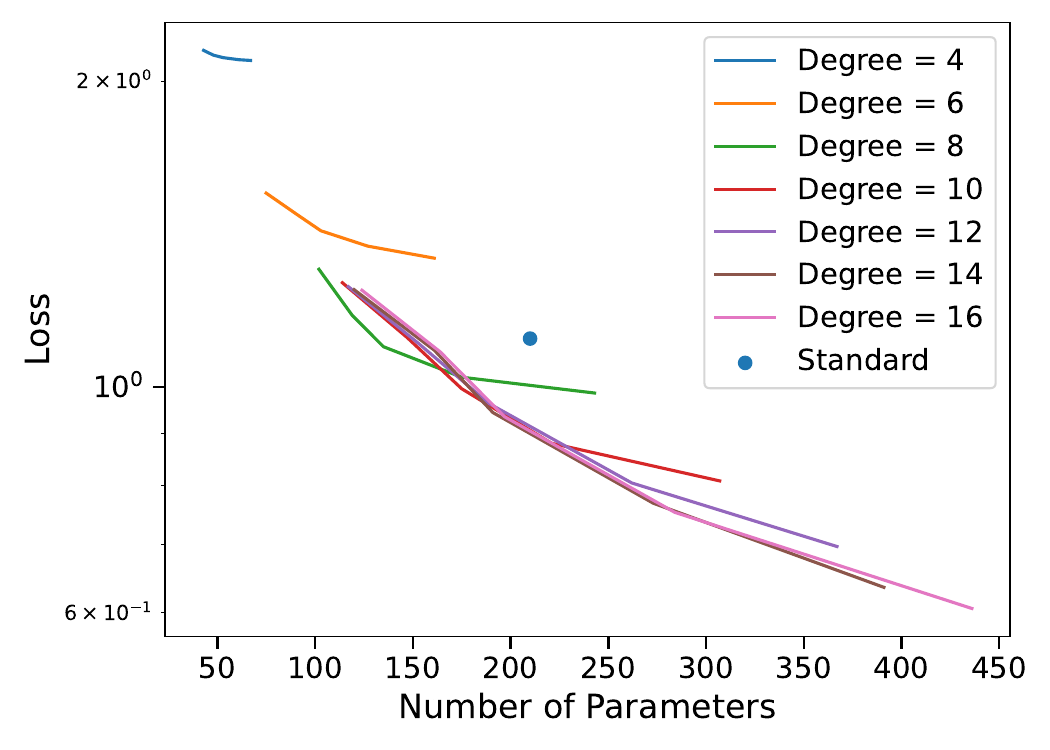}
\caption{Loss versus number of parameters for different polynomial degrees. The regularization parameter was varied in the lasso regression to yield a Pareto front in model accuracy and complexity for each degree.}
\label{fig:4}
\vspace{-5mm}
\end{figure}

\begin{figure}[t]
\captionsetup[subfigure]{labelformat=empty}
  \centering
  \subfloat[][(a)]{\includegraphics[width=.46\textwidth]{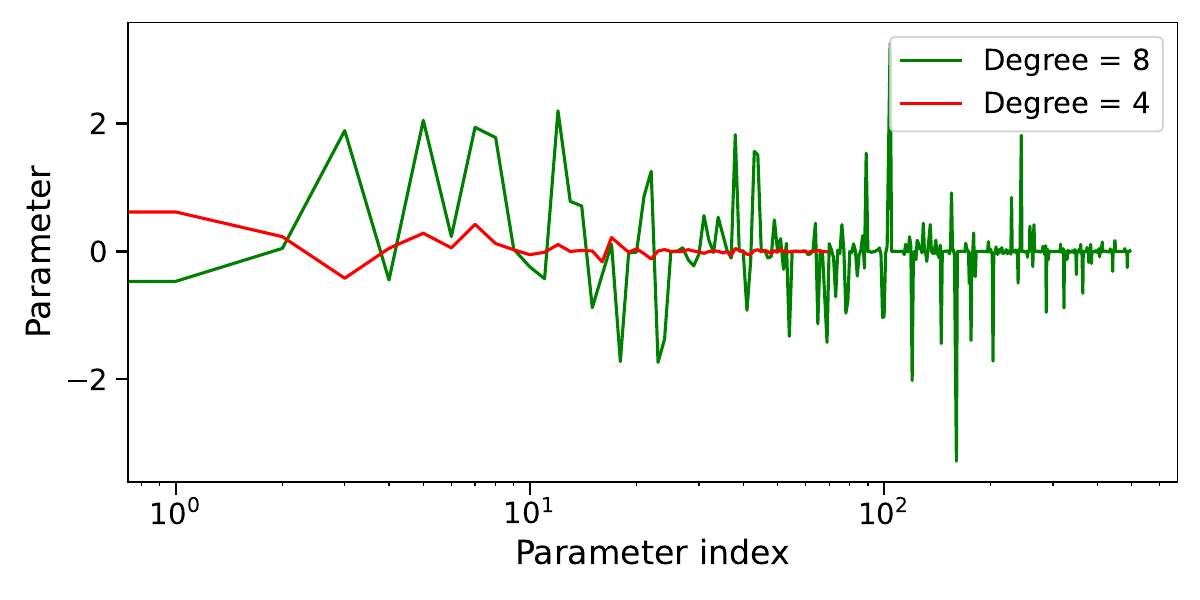}}\quad 
  \subfloat[][(b)]{\includegraphics[width=.46\textwidth]{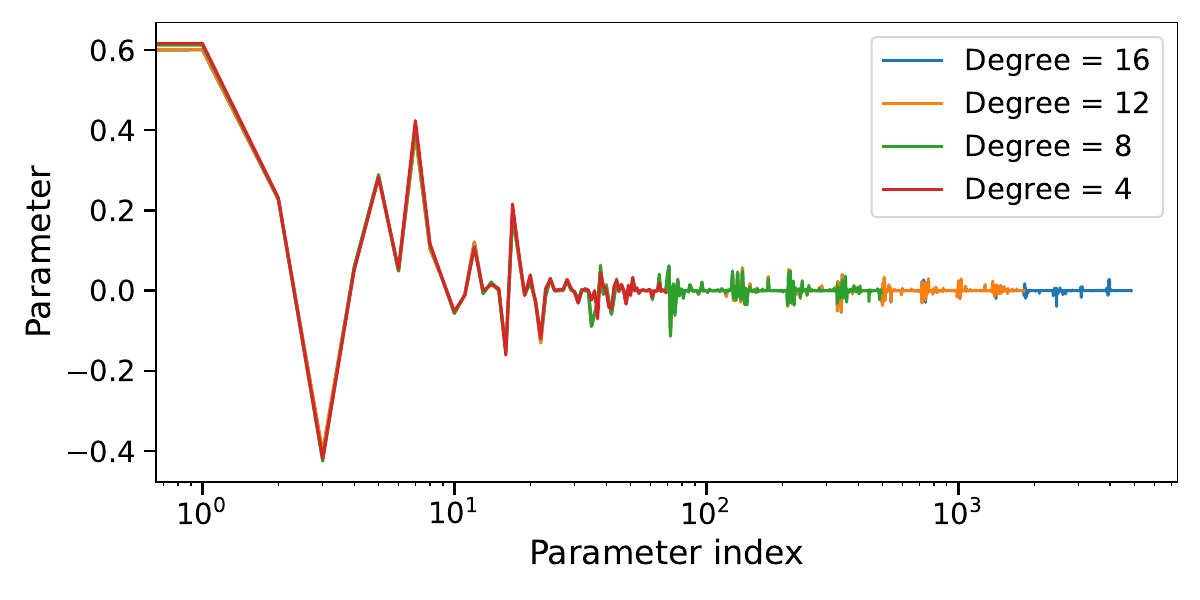}}\\
\subfloat[][(c)]{\includegraphics[width=\textwidth, height=5cm]{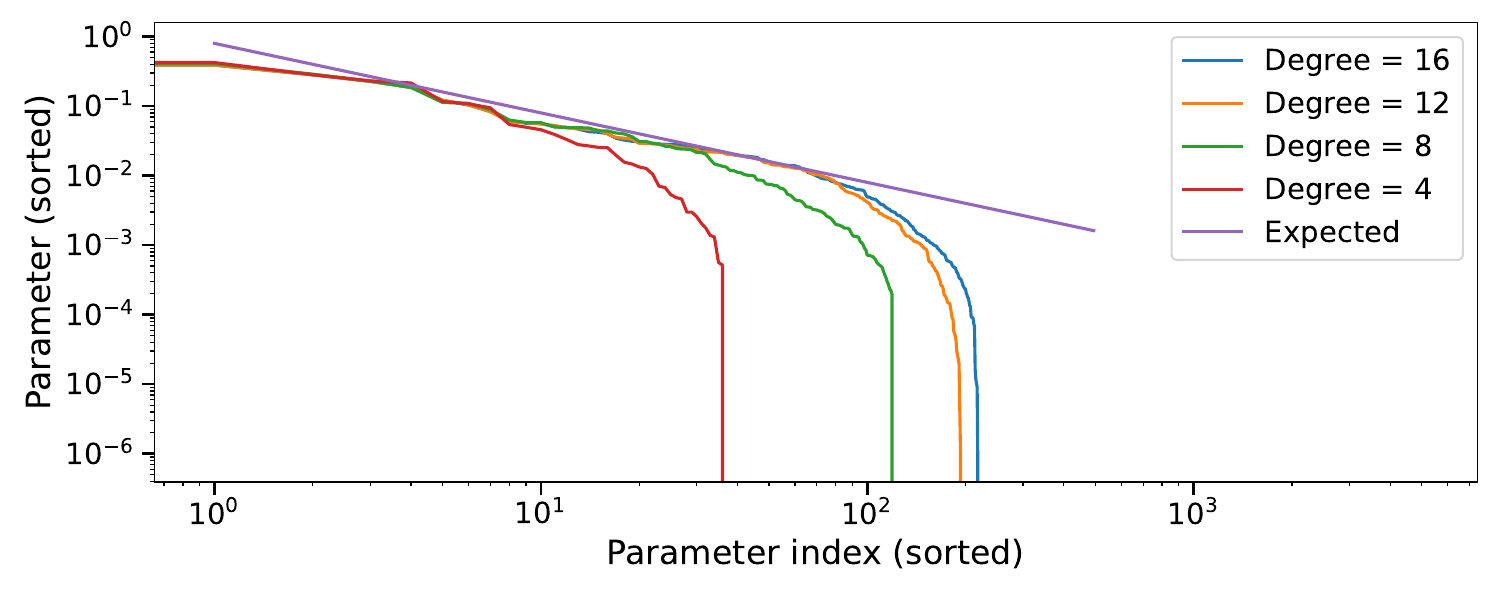}}
\caption{Parameters (or polynomial coefficients) and how they change for different polynomial degrees for (a) simple regression and (b) sparse regression (using Legendre basis). (c) Sorted sparse‐regression coefficients (Legendre basis) versus parameter index on a logarithmic x–axis show a clear, Fourier–like decay with order—approximately $1/n$—that is stable across model capacities (degrees 4, 8, 12, 16), indicating a hierarchical structure where lower–order terms dominate and higher–order terms provide incremental refinement.}
    \label{fig:5}
\vspace{-5mm}
\end{figure}

To make the fitted model class explicit, let
$\tilde{T}_a$, $\tilde{v}_a$, $\widetilde{\Delta T_r}$, and $\widetilde{rH}$
denote the normalized versions of $T_a$, $v_a$, $T_r-T_a$, and $rH$,
respectively, each mapped to the interval $[-1,1]$ according to the ranges in
Table~\ref{tab:vars}. In this formulation, relative humidity is retained as an
input variable in order to account for the effect of water vapor. The
approximation of the UTCI offset can then be written in the general form
\begin{equation}
\widehat{\mathrm{Offset}}(T_a,v_a,T_r-T_a,rH)
=
\sum_{\boldsymbol{\alpha}\in\mathcal{A}_p}
c_{\boldsymbol{\alpha}}
\prod_{j=1}^{4} P_{\alpha_j}(x_j),
\label{eq:legendre_expansion}
\end{equation}
where $P_n(\cdot)$ denotes the Legendre polynomial of degree $n$,
$(x_1,x_2,x_3,x_4)=
(\tilde{T}_a,\tilde{v}_a,\widetilde{\Delta T_r},\widetilde{rH})$,
$\boldsymbol{\alpha}=(\alpha_1,\alpha_2,\alpha_3,\alpha_4)$ is a multi-index,
and
\[
\mathcal{A}_p=
\left\{
\boldsymbol{\alpha}\in\mathbb{N}_0^4:
\alpha_1+\alpha_2+\alpha_3+\alpha_4\le p
\right\}
\]
is the set of all basis terms up to total polynomial degree $p$. Thus, the
model is a linear combination of products of Legendre polynomials in the four
normalized environmental variables. For a given maximum degree $p$, the full
candidate basis contains $\binom{p+4}{4}$ terms, which yields the sequence
$70,210,495,\ldots$ reported in Table~\ref{tab:1} for degrees $4,6,8,\ldots$.
Sparse orthogonal regression restricts this expansion by retaining only a subset
of the candidate terms,
\begin{equation}
\widehat{\mathrm{Offset}}(T_a,v_a,T_r-T_a,rH)
=
\sum_{\boldsymbol{\alpha}\in S_p}
c_{\boldsymbol{\alpha}}
\prod_{j=1}^{4} P_{\alpha_j}(x_j),
\label{eq:sparse_legendre_expansion}
\end{equation}
where $S_p\subseteq\mathcal{A}_p$ is selected by the Lasso regularization. The
number of active parameters therefore depends on two factors: the maximum
polynomial degree, which determines the size of the candidate pool, and the
regularization strength, which determines how many of those candidate terms are
retained in the final model. This is the reason why the number of parameters
changes across polynomial degrees and also along the Pareto fronts shown in
Fig.~\ref{fig:4}. In this sense, the approximation can be viewed as a
Fourier-like decomposition in an orthogonal polynomial basis, where lower-order
terms capture the dominant structure of the UTCI offset and higher-order terms
provide progressively finer corrections. A key advantage of the orthogonal basis
is that it yields order-by-order consistency, see Fig.~\ref{fig:5}: when higher-degree terms are
introduced, the coefficients associated with lower-order structure remain much
more stable than in regressions based on ordinary monomials.

Linear regression without any sparsity constraints shows improved performance at higher degrees, with test loss reducing as model capacity increases. However, this comes with a dramatic increase in the number of parameters; it reaches over 1800 coefficients by degree 12. Furthermore, the discrepancy between train and test losses at higher degrees (e.g., 0.62$^{\circ}$C vs.~0.54$^{\circ}$C at degree 12) indicates overfitting, despite the improved predictive accuracy. The resulting models are also substantially more complex, raising concerns regarding interpretation and generalization. Sparse regression with standard polynomial bases shows similar performance at low degrees but fails to converge beyond the 6th degree. This indicates that enforcing sparsity in a poorly conditioned basis becomes increasingly difficult as model complexity grows.

In contrast, sparse regression using an orthogonal Legendre basis (or sparse orthogonal regression) exhibits superior stability and accuracy across all degrees. It outperforms the baseline 6th-degree polynomial fit from degree 8th onward, achieving a test loss of 0.88$^{\circ}$C at degree 10 with only 209 parameters—almost the same count as the original benchmark model, but with improved generalization. As the degree increases to 16, the loss reduces further to 0.60$^{\circ}$C using 424 parameters—a fraction of those used by the corresponding standard regression model. The orthogonality of the Legendre basis likely contributes to better numerical conditioning, facilitating sparse model discovery even at high degrees. These results emphasize the importance of basis selection and regularization strategy in symbolic regression tasks. Sparse methods, when combined with well-structured bases like Legendre polynomials, offer a promising path toward accurate, compact, and interpretable models in high-dimensional settings.

Furthermore, optimization of nonlinear objective functions using gradient-based algorithms can be computationally intensive, especially in high-dimensional spaces where convergence is slow and local minima may hinder performance. In contrast, the regression-based approach proposed in this article—particularly through sparse regression with orthogonal polynomials—offers significantly faster computation. By framing the problem as a structured regression task rather than a nonlinear optimization, the method avoids costly iterative procedures and scales efficiently with dimensionality, making it highly suitable for rapid modeling of complex environmental indices like the UTCI.

Fig.~\ref{fig:4} illustrates the relationship between model complexity (measured by the number of parameters) and prediction accuracy (log-scaled loss) for sparse regression models using Legendre polynomial bases of varying degrees. Each curve corresponds to a fixed polynomial degree, ranging from 4 to 16, with points reflecting models of increasing complexity obtained through regularization. A clear trend is observed: for a given polynomial degree, increasing the number of parameters generally results in improved model accuracy (i.e., lower loss). However, diminishing returns set in, and the rate of improvement flattens. More notably, the envelope formed by the lowest loss at each level of complexity across all degrees traces an emergent Pareto front \citep{smits2005pareto}. This front captures the trade-off between model simplicity and predictive performance.

Higher-degree models (e.g., degrees 12–16) dominate this frontier at higher parameter counts, offering better loss with only marginal increases in complexity. In contrast, lower-degree models saturate quickly, highlighting their limited expressivity. The Pareto front thus reflects the optimal set of models that balance accuracy and sparsity, guiding model selection under complexity constraints. The use of Legendre polynomials ensures numerical stability and encourages efficient basis representations, which supports the recovery of compact yet accurate models in this sparse regression setting.

In Fig.~\ref{fig:5}(a) and (b) we visualize the behavior of regression coefficients obtained from simple regression and sparse regression with orthogonal Legendre polynomials. Both plots use a logarithmic x-axis to indicate the parameter index and reveal how coefficients evolve as higher-degree polynomial terms are introduced. In Fig.~\ref{fig:5}(a), each line corresponds to simple regression solutions using polynomial bases of increasing degree. The x-axis denotes the index of polynomial terms (sorted or sequential), while the y-axis shows the corresponding coefficient values. A key observation is that the coefficients of lower-degree terms (left side of the plot) are not stable across model orders. As higher-degree terms are added, previously estimated lower-order coefficients shift significantly, often changing sign and magnitude. 

Figure~\ref{fig:5}(b) presents coefficient values for sparse regression using Legendre polynomials, with colors indicating contributions from different polynomial degrees. Here, a contrasting pattern emerges: coefficients associated with lower-degree terms remain stable as higher-degree terms are added. New coefficients primarily emerge in the higher-order region of the x-axis, without disturbing the existing ones. This stability results from the orthogonality of the Legendre basis, which decorrelates the polynomial terms and enables additive refinement without re-tuning existing coefficients.

The contrast between the Figs.~\ref{fig:5}(a) and (b) underscores the advantage of orthogonal polynomial bases in sparse regression. Simple regression results in unstable, entangled coefficient estimates that shift with basis expansion, complicating interpretability and reuse. Sparse regression with ordinary polynomial bases fails to converge for higher degrees. In contrast, sparsity and orthogonal polynomials yield stable, hierarchical models where lower-order structure is preserved and higher-order terms incrementally enrich the representation. This behavior is particularly valuable for symbolic regression and interpretable modeling, where each term ideally reflects a distinct, meaningful contribution to the model output.

In Fourier analysis, the magnitude of coefficients typically decays as $1/n$ (where $n$ is the order of the term) for functions of bounded variation \citep{stein2011fourier} -- a class that includes many naturally occurring signals and is a reasonable assumption for observational data. This decay reflects the fact that higher-order (or higher-frequency) components contribute less to the overall structure of such functions. A similar trend is observed in sparse regression using orthogonal polynomial bases, see Fig.~\ref{fig:5}(c). When coefficients are sorted by magnitude, they exhibit a clear decreasing pattern, analogous to the Fourier case, with lower-order terms capturing the dominant structure and higher-order terms refining the approximation in a controlled manner.

This suggests that through the use of sparse regression with an orthogonal polynomial basis, we have achieved a Fourier-like decomposition of the UTCI Offset in the Legendre basis (instead of the trigonometric one). This has a number of theoretical advantages: due to the orthogonality of the basis functions, the decomposition minimizes the $L_{2}$ distance (least squares) between approximation and function, guaranteeing the best possible polynomial fit for a given model complexity \citep{stein2011fourier}. Additionally, the coefficients are uncorrelated and hierarchically structured, ensuring that lower-order components remain stable as higher-order terms are added—enhancing both interpretability and numerical robustness. 

Based on the analysis results and one of the initial goals (that the new approximation should have comparable computational complexity to the existing one), we selected the sparse regression model based on tenth-degree Legendre polynomials as the most suitable approximation. The final version of the new polynomial, which has 209 coefficients, was calculated using the whole dataset of tabulated values.

\begin{figure}[t]
\captionsetup[subfigure]{labelformat=empty}
  \centering
  \subfloat[][(a)]{\includegraphics[width=0.67\textwidth, height=5cm]{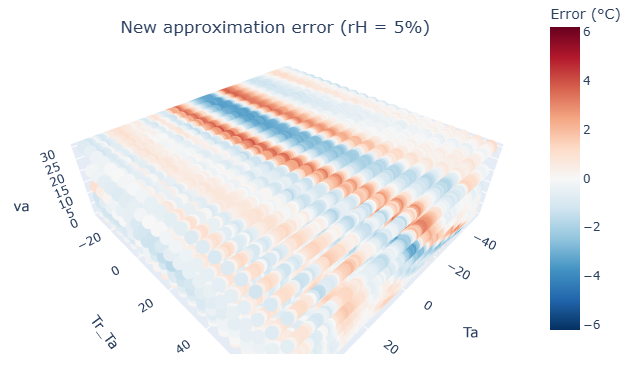}}\\
  \subfloat[][(b)]
  {\includegraphics[width=.48\textwidth]{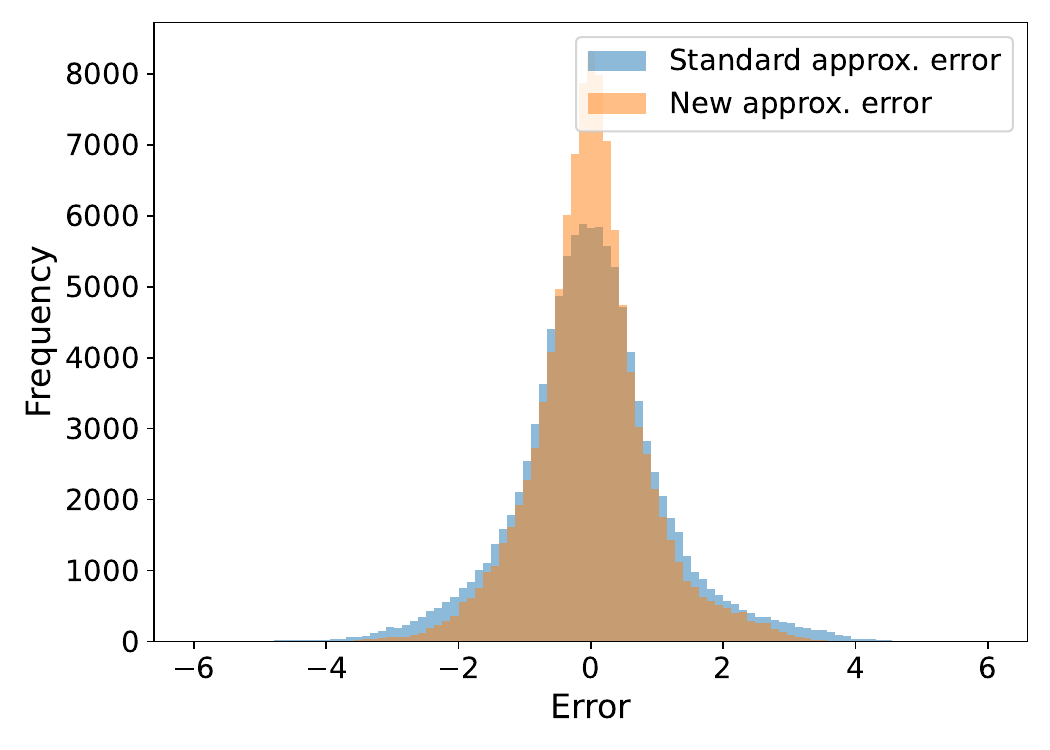}}\quad
  \subfloat[][(c)]{\includegraphics[width=.48\textwidth]{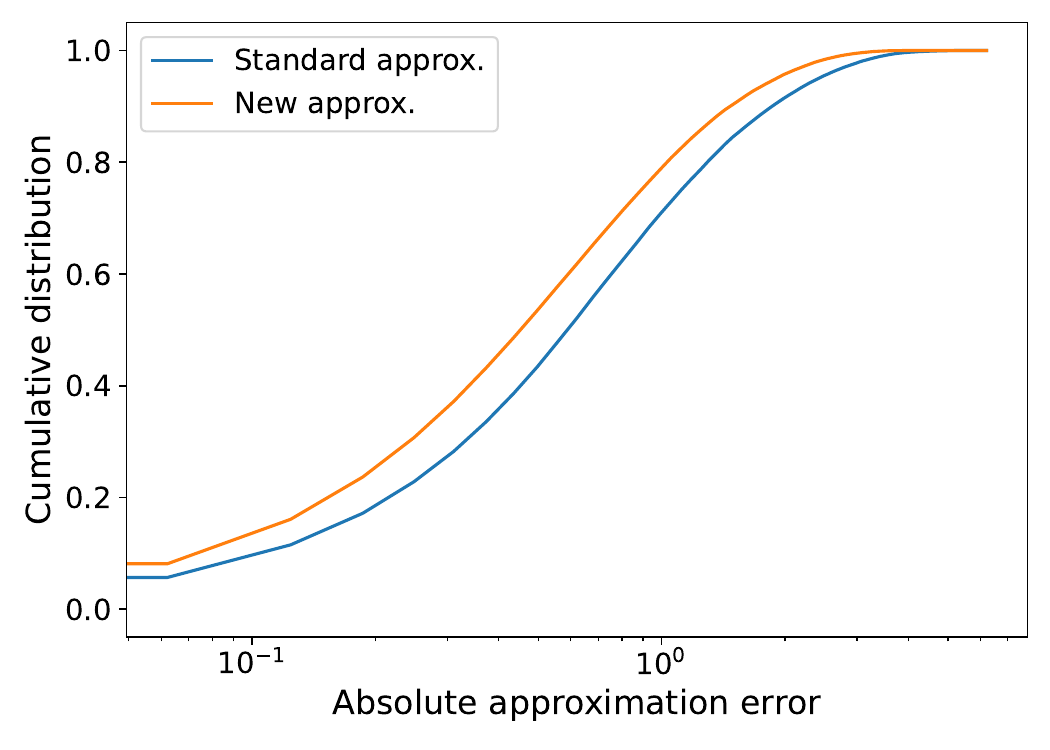}}
    \caption{(a) Spatial distribution of the UTCI Offset error (approximation minus reference) for the new sparse-model-based approximation at a fixed relative humidity of 5\%, showing small, smoothly varying discrepancies. (b) Comparison of error histograms for the standard UTCI approximation and the new approximation based on the tenth-degree Legendre polynomials. (c) Cumulative distributions of the absolute errors of the two approximations.}
    \label{fig:7}
\vspace{-5mm}
\end{figure}

Fig.~\ref{fig:7}(a) shows the spatial distribution of the Offset errors for the new approximation at a fixed relative humidity of 5\%. The errors are small and smoothly varying, indicating good agreement across the input space. Fig.~\ref{fig:7}(b) presents a comparison of error histograms for both the standard and new approximations. The sparse-model-based approximation produces a narrower, more sharply peaked distribution centered at zero, highlighting a reduction in error variance and suggesting better generalization. Fig.~\ref{fig:7}(c) shows the cumulative distribution of absolute errors for the two approximations. The curve for the new approximation rises more steeply and reaches higher cumulative values at lower error thresholds, indicating that a larger proportion of predictions fall within smaller error margins. 

Table~\ref{tab:2} summarizes the most relevant properties of the two approximations. The results show a clear improvement in accuracy: the new approximation not only substantially reduces the average errors (i.e, the mean error, the mean absolute error, and the root mean square error) but also drastically reduces the frequency of large deviations compared to the standard approximation. For example, the frequency of absolute errors larger than $2^{\circ}$C is halved from 8\% to 4\%, the frequency of errors larger than $3^{\circ}$C reduces from 2\% to 0.5\%, while the frequency of errors larger than $4^{\circ}$C reduces from 0.3\% to 0.01\%. These results clearly show the added benefits of the new approximation and confirm that the sparse regression approach can achieve comparable or improved predictive accuracy while maintaining interpretability and model parsimony.

\begin{table}[t]
\centering
\begin{tabular}{l|cc}
 & Standard  & New  \\
 &  approximation &  approximation \\ \hline
Polynomial degree & 6th & 10th \\
Basis functions & monomials & Legendre \\
Number of coefficients & 210 & 209 \\
Mean Error & $1.7\cdot10^{-3}$ $^{\circ}$C (0.35 $^{\circ}$C) & $-2.7\cdot10^{-15}$ $^{\circ}$C (0.22 $^{\circ}$C)  \\
Mean Absolute Error & 0.81 $^{\circ}$C (1.33 $^{\circ}$C) & 0.64 $^{\circ}$C (0.71 $^{\circ}$C) \\
Root Mean Square Error & 1.17 $^{\circ}$C (2.77 $^{\circ}$C) & 0.88 $^{\circ}$C (0.96 $^{\circ}$C) \\
Freq. of abs. errors larger than 2$^{\circ}$C & 8.4 \% (15.5 \%)& 4.2 \% (5.0 \%)\\
Freq. of abs. errors larger than 3$^{\circ}$C & 2.2 \% (6.3 \%)& 0.50 \% (0.60 \%)\\
Freq. of abs. errors larger than 4$^{\circ}$C & 0.34 \% (3.8 \%)& 0.011 \% (0.10 \%)\\
Freq. of abs. errors larger than 5$^{\circ}$C & 0.038 \% (3.3 \%)& 0.00096 \% (0 \%)\\
\end{tabular}
\caption{Comparison of properties of the standard \citep{Brode2012} and new polynomial approximations of UTCI Offset function. The values outside of the parentheses reflect the evaluation of the approximations on the full dataset of $104\,643$ accurate Offset values provided by \cite{Brode2012}. The values shown in the parentheses reflect the evaluation using the independent dataset of 1000 accurate UTCI values \citep{ZenodoBrode2021}, which were not used during the development of the new approximation. Both approximations are only valid for the intervals of environmental variables available in the full dataset (Table~\ref{tab:vars}).}
\label{tab:2}
\end{table}

We also evaluated the new approximation on the independent dataset of 1000 accurate UTCI values, which were not used during the development of the approximation. This dataset was prepared by the authors of the \cite{Brode2012} paper, and is freely available on a Zenodo repository \citep{ZenodoBrode2021}. Similarly to the evaluation of the new approximation on the full dataset, evaluation on the independent dataset shows a substantial reduction of the mean errors and a drastic reduction in the frequency of large errors compared to the standard approximation (Table~\ref{tab:2}). 

Since the new approximation was determined using the full dataset of accurate Offset values \citep{Brode2012}, it is, same as the standard approximation, only valid for the intervals of environmental variables available in this dataset (Table~\ref{tab:vars}). Using the approximation for conditions outside of these intervals can potentially lead to large errors or unrealistic results and should be avoided.

\section{Conclusions}

The goal of this study was to develop an improved version of the polynomial approximation -- one that would have comparable computational complexity to the existing approximation but would be more robust in terms of numerical stability and substantially more accurate, particularly in reducing the frequency of larger errors. This goal was successfully achieved using sparse regression with an orthogonal polynomial basis.

Sparse regression methods, such as LASSO, helped reduce overfitting and improve interpretability. As we have shown, the choice of basis functions is crucial: orthogonal polynomials like Legendre polynomials offer better numerical stability and conditioning than monomials. They enable hierarchical models where higher-order terms don’t affect lower-order estimates, making them especially useful in sparse, interpretable models. Empirical results support these theoretical advantages. 

Using sparse regression with an orthogonal polynomial basis (or sparse orthogonal regression), we have: 
\begin{enumerate}[label=(\alph*)]
    \item Achieved substantially better accuracy -- compared to the standard approximation, the new approximation not only substantially reduces the average errors (i.e, the mean error, the mean absolute error, and the root mean square error) but also drastically reduces the frequency of large errors. 
    \item Retained a comparable computational complexity -- the number of coefficients is almost the same for both approximations, meaning the computational complexity is comparable.
    \item  Found a Pareto front for different model complexities -- loss curves reveal that sparse models with orthogonal bases efficiently populate a Pareto front, balancing complexity and accuracy.
    \item Determined coefficients consistent over models with different capacities - coefficient plots for models built on orthogonal bases show the progressive inclusion of higher-order components without disrupting lower-order structure, in contrast to models using simple regression and ordinary polynomials.
    \item Achieved successful generalization -- training the model over only 20\% of the data, while testing was performed over the other 80\%, highlights successful generalization. The results are also robust under bootstrapping.
    \item Essentially decomposed the UTCI in a Fourier expansion with a Legendre-polynomial basis, with parameters scaling as expected. Thus, we are arguably close to the theoretical optimum results for a robust approximation in the $L_{2}$ metric (or least squares).
\end{enumerate}

Sparse orthogonal regression provides an effective framework for constructing accurate and numerically stable polynomial approximations of the UTCI. Our main contribution is therefore not methodological novelty in sparse regression itself, but the use of an orthogonal polynomial basis as a practical approximation strategy with favorable numerical properties, including order-by-order consistency and stable low-order truncations. In addition, the results obtained from random train--test splits, together with their robustness under bootstrapping, show that using only 20\% of the data for training is not a requirement of the method, but a deliberately stringent test of generalization. The comparable performance on the remaining 80\% of the data indicates that the approach remains accurate, robust, and efficient even under a severe limitation in the number of training data points, while remaining well suited for practical applications that require portability and ease of implementation.

We have also prepared an easy-to-use Python function for the new approximation (please refer to the Code and data availability section on how to obtain the code). The code relies only on basic mathematical operations, which makes it easy to adapt to other programming languages, such as Fortran or C++. We also implemented a check to see if the environmental state falls within the domain of validity of the approximation. If this is not the case, the code produces a warning that the resulting UTCI values could have large errors or be unrealistic.

\section*{Code and data availability}

The code used to calculate the new UTCI approximation, generate the reported model comparisons, and reproduce the analysis, tables, and figures presented in this paper is archived on Zenodo \citep{ZenodoRoman2025}. The archive includes reproducibility instructions and the required Python environment specification.

The offset data used for fitting and evaluating the approximation are the supplementary material of \cite{Brode2012}, available from the publisher as electronic supplementary material and downloaded automatically by the reproduction code. The independent UTCI test data used for additional validation are publicly available on Zenodo \citep{ZenodoBrode2021}. No additional non-public data were used.

\section*{Author Contributions}

S.R. - Conceptualization, Data curation, Formal analysis, Investigation, Methodology, Software, Validation, Visualization, Writing (original draft preparation), G.S. - Conceptualization, Resources, Validation, Software, Writing (review and editing), L.T. - Conceptualization, Methodology, Project administration, Supervision, Validation, Writing (review and editing), S.D. - Conceptualization, Funding acquisition, Project administration, Resources, Supervision, Writing (review and editing).

\section*{Conflict of Interest Statement}
The authors declare no conflicts of interest.

\section*{Disclaimer}
Co-funded by the European Union. Views and opinions expressed are however those of the author(s) only and do not necessarily reflect those of the European Union or European Research Executive Agency. Neither the European Union nor the granting authority can be held responsible for them.

\section*{Acknowledgments}
The authors appreciate the fruitful discussions within the SHED discussion group with Jure Brence, Nina Omejc, Sebastian Me\v{z}nar, and Bo\v{s}tjan Gec.

\section*{Funding}
This publication is supported by the European Union's Horizon Europe research and innovation programme under the Marie Sk\l{}odowska-Curie Postdoctoral Fellowship Programme, SMASH co-funded under the grant agreement No.~101081355. The operation (SMASH project) is co-funded by the Republic of Slovenia and the European Union from the European Regional Development Fund. The authors acknowledge the financial support of the Slovenian Research Agency via the Gravity project \textsl{AI for Science}, GC-0001 and of the Slovenian Research And Innovation Agency (research core funding No.~P1-0188).

%
%
%
%
\bibliographystyle{Copernicus}
\bibliography{mybib}

\end{document}